\documentclass[%
 reprint,
 onecolumn,
 amsmath,amssymb,
 aps,
 prl
]{revtex4-2}

\usepackage{amsmath}

\begin{document}
\title{Comment on "Universal Bound on Energy Cost of Bit Reset in Finite Time"}
\author{Jie Gu }
\affiliation{Chengdu Research Institute of Education Science, Chengdu, Sichuan 610000, China}
\author{X.G. Zhang}
\affiliation{Department of Physics, Center for Molecular Magnetic Quantum Materials and Quantum Theory Project,
University of Florida, Gainesville, Florida 32611, USA}
\date{\today}

\maketitle

In Ref. \cite{Zhen2021}, as a crucial step to their main result Eq. (5), the authors proved an important identity (Eq. (6) therein),
\begin{equation}
\label{eq:cl}
    W_{\text{pn}}  = T \left[ \Delta S_1(p \| \gamma)+ \Sigma \right],
\end{equation}
where $k_B=1$, $\Delta S_1 = \sum_i p_i \ln(p_i/\gamma_i)$ is the relative entropy, and $\gamma$ the thermal state. This formula points out that, the work penalty (also called dissipated work in, e.g., \cite{miller2019}), which is the irreversible work on top of the quasi-static work due to finite-time process, consists of two contributions: the change in the relative entropy with respect to the thermal states and the entropy production.
As detailed in its Supplemental Material, Eq. \eqref{eq:cl} was proved by decomposing $\dot S_1(p(t)\| \gamma(t))$ into two terms and identifying each term as $\dot \Sigma$ and $\beta \dot W_\text{pn}$ respectively. Here we present a quantum-mechanical generalization of Eq. \eqref{eq:cl} with an alternative derivation.

Consider an open quantum system interacting  with a heat bath. The two end points, i.e., the density matrices \(\rho(0)\) and \(\rho(\tau)\), are fixed.
The
average work performed on the system associated with the quasi-static protocol  is
\begin{equation}
\label{eq:efe}
  W_\text{qs} = \Delta F,  
\end{equation}
where \(F\) is the equilibrium free energy, given
by \(F:= -T \ln Z\), and \(Z\) the partition function
\(Z: = \text{Tr}(-e^{\beta \hat H})\). The work associated to a generic driving
protocol is
\begin{equation}
  W = \Delta E - Q\\
= \Delta E - T(\Delta S_1-\Sigma) \\
= \Delta F_1 + T \Sigma  ,
\end{equation}
where we have used the first law of thermodynamics, the definition of
entropy production \(\Sigma=\Delta S_1-Q/T\), and the definition of nonequilibrium
1-free energy \(F_1=E-TS_1\). The work penalty  is then
\begin{equation}
  W_\text{pn} := W-W_\text{qs}  = T \left[ \Delta S_1(\rho||\rho^ \text{G})+ \Sigma \right],  
\end{equation}
where we have plugged in Eq. \eqref{eq:efe}  and rewritten the nonequilibrium free energy as
\(F_1 = T S_1(\rho \| \rho^\text{G} )+F\) (see, e.g., \cite{Sagawa2020}). Here $\rho^\text{G}$ denotes the Gibbs state. This equation is universal and applicable in quantum processes. By taking only diagonal elements in the energy eigenbasis of the Hamiltonian, the equation above reduces to its classical counterpart, Eq. \eqref{eq:cl}.

\bibliographystyle{unsrt}

\bibliography{ref}

\end{document}